\pdfoutput=1
\documentclass[12pt]{article}

\usepackage[T1]{fontenc}
\usepackage[utf8]{inputenc}
\usepackage[english]{babel}

\usepackage{graphicx}
\usepackage{subcaption}
\usepackage{amssymb}
\usepackage{amsmath}
\usepackage{amsthm}
\usepackage{bm}
\usepackage{mathtools}
\usepackage{booktabs}
\usepackage{setspace}
\usepackage{xspace}
\usepackage{siunitx}
\usepackage[switch]{lineno}
\usepackage{xurl}

\usepackage[colorlinks=true,linkcolor=black,citecolor=blue,urlcolor=blue]{hyperref}
\usepackage{varioref}
\usepackage{cleveref}

\newcommand{\arco}[1]{Ar/CO$_{2}$}
\newcommand{\litium}[1]{\ensuremath{{}^{#1}\mathrm{Li}}}
\newcommand{\boron}[1]{\ensuremath{{}^{#1}\mathrm{B}}}
\newcommand{\helium}[1]{\ensuremath{{}^{#1}\mathrm{He}}}

\newcommand{\GEANT}{{\textsc{Geant4}}\xspace}

\DeclareCaptionFormat{custom}{\textbf{#1#2}#3}
\title{Cold Neutron Imaging and Efficiency Measurements with a Boron-10 Coated Double-GEM Detector}

\author{%
\parbox{0.96\textwidth}{\centering
WooJong Kim\textsuperscript{1,*},
DongHyun Kim\textsuperscript{1},
Minjae Kwon\textsuperscript{3},
Jason Sang Hun Lee\textsuperscript{1},
Hyupwoo Lee\textsuperscript{1},
Inkyu Park\textsuperscript{1},
Donghyun Song\textsuperscript{1},
Inseok Yoon\textsuperscript{2},
Myeonghun Choi\textsuperscript{1}
\\[0.8em]
\small
\textsuperscript{1}Department of Physics, University of Seoul, Republic of Korea\\
\textsuperscript{2}Institute of Basic Sciences, Seoul National University, Republic of Korea\\
\textsuperscript{3}Department of Physics, University of Turin, Italy\\[0.6em]
\textsuperscript{*}Corresponding author: \href{mailto:kwj1180@uos.ac.kr}{kwj1180@uos.ac.kr}
}%
}

\date{}

\begin{document}
\setlength{\baselineskip}{17pt}

\maketitle

\begin{abstract}
A \boron{10}-coated double-GEM neutron detector (BGEM) was developed as a \helium{3}-free cold-neutron beamline detector using a single \(\mathrm{B_4C}\) converter cathode and a 512-channel APV25 orthogonal-strip readout over \qtyproduct[product-units=single]{10 x 10}{cm^2}.
The detector was tested at the HANARO Bio-REF beamline with a monochromatic \qty{4.5}{\angstrom} beam (\(E_n=\qty{4.03}{meV}\)).
The absolute detection efficiency relative to a \litium{6}-based Ce:LiCAF reference detector was \(\varepsilon_{\mathrm{BGEM}}=\qty[separate-uncertainty=true]{8.69(20)}{\percent}~\mathrm{(stat.)}\).
The pulse-height spectrum was qualitatively consistent with \GEANT energy-deposition simulations, and Cd-mask imaging yielded a Gaussian-equivalent edge-spread width of \(\sigma = 555 \oplus 102~\mathrm{\mu m}\).
These results establish a cold-neutron beamline benchmark for a single-converter BGEM detector with full-strip APV25 readout.
\end{abstract}


\newpage

\section{Introduction}
\label{sec:intro}

Neutron detection technology plays a pivotal role in diverse applications, ranging from neutron diagnostics in nuclear fusion and boron neutron capture therapy (BNCT)~\cite{JRC48612} to process monitoring in nuclear fuel cycle technologies, such as pyroprocessing~\cite{Coble01122020}.
Consequently, the demand for cost-effective and high-performance neutron detection systems is steadily increasing across these domains.

Since neutrons carry no electric charge, their detection relies on secondary charged particles generated via scattering or nuclear capture reactions.
Unlike direct ionization detectors used for charged particles or gamma rays, neutron detectors require a specific neutron-reactive converter material.
Historically, standard detectors have employed \({}^{3}\mathrm{He}\) or, more recently, \({}^{6}\mathrm{Li}\) as converter materials.
While \({}^{3}\mathrm{He}\) gas offers high detection efficiency, it is a strategic resource facing severe supply shortages~\cite{KOUZES20101035}.
Alternatively, \({}^{6}\mathrm{Li}\) is a viable option; however, the intrinsic properties of its crystal form present significant challenges for large-area detector fabrication.
Consequently, we investigated alternative materials with high neutron capture cross-sections to develop a robust and scalable neutron detector.

Materials such as \({}^{157}\mathrm{Gd}\) and \({}^{10}\mathrm{B}\) are recognized as viable alternatives to \({}^{3}\mathrm{He}\) due to their large neutron capture cross-sections~\cite{BROWN20181}.
Although \({}^{157}\mathrm{Gd}\) exhibits a high capture probability, it emits low-energy internal conversion electrons, which complicates the discrimination of neutron signals from gamma-ray backgrounds~\cite{firestone2004database}.
Therefore, this study utilizes \({}^{10}\mathrm{B}\) due to its clearer signal characteristics.
As illustrated in Eq.~\ref{eq:boron_capture}, upon neutron absorption, \({}^{10}\mathrm{B}\) undergoes the \({}^{10}\mathrm{B}(n,\alpha){}^{7}\mathrm{Li}\) reaction, producing energetic alpha and lithium ions, often accompanied by gamma emission~\cite{international2023iaea}.
The high Q-value of this reaction ensures the generation of large ionization signals, facilitating robust neutron detection.
Furthermore, the cost-effectiveness of boron supports the scalability required for manufacturing large-area detectors.

\begin{equation}
\label{eq:boron_capture}
\begin{split}
    {}^{10}\mathrm{B} + n
    &\xrightarrow{\phantom{0}6.3\%\phantom{0}} \alpha\,(1.78~\mathrm{MeV}) + {}^{7}\mathrm{Li}\,(1.02~\mathrm{MeV}) \\
    &\xrightarrow{93.7\%\phantom{0}} \alpha\,(1.47~\mathrm{MeV}) + {}^{7}\mathrm{Li}^{*} \\
    &\phantom{\xrightarrow{93.7\%} \alpha\,(1.47~\mathrm{MeV}) + {}} \hookrightarrow {}^{7}\mathrm{Li}\,(0.84~\mathrm{MeV}) + \gamma\,(0.48~\mathrm{MeV}) .
\end{split}
\end{equation}

To realize a scalable neutron detector, we adopted Gas Electron Multiplier (GEM) technology~\cite{NIMA1997531}, a mature platform for large-area and position-sensitive gaseous detectors.

Boron-converter GEM neutron detectors have previously been demonstrated in several forms, including CASCADE-type multilayer detectors~\cite{Klein2011,Kohli2016}, high-efficiency and sealed GEM-based detectors for neutron beamlines~\cite{Zhou2020,Zhou2021}, and a double-GEM detector with a \({}^{10}\mathrm{B}_{4}\mathrm{C}\)-coated cathode using resistive charge-division readout~\cite{SerraFilho2022}. These works established the feasibility of boron-coated GEM detectors for thermal-neutron detection.

The present work is positioned not as a new neutron-conversion concept, but as a cold-neutron beamline benchmark of a single-converter cathode-type double-GEM detector with full 512-channel APV25 strip readout. By preserving strip-level charge information event by event, the detector allows the absolute cold-neutron detection efficiency, pulse-height consistency with \GEANT simulations, and reconstructed imaging performance to be assessed within the same detector architecture.

\section{Detector Design and Fabrication}
\label{sec:design}

\subsection{Geometry Optimization via Geant4 Simulation}
\label{subsec:geant4_optimization}

In this study, \GEANT~\cite{ALLISON2016186} simulations were performed to optimize the detector geometry and validate the feasibility of the selected detector configuration.
First, we evaluated various coating configurations for the boron-10 carbide (\({}^{10}\mathrm{B}_4\mathrm{C}\)) converter.
While coating directly on the GEM foils was considered to potentially enhance detection efficiency, it was excluded from the final design due to concerns regarding high-voltage stability.
Consequently, a configuration utilizing a \({}^{10}\mathrm{B}_4\mathrm{C}\)-coated cathode was selected as the baseline geometry.

\GEANT was used as a unified detector-response framework to describe reaction-product transport, energy deposition in the drift volume, detection efficiency, and the expected pulse-height spectrum from the \({}^{10}\mathrm{B}(n,\alpha){}^{7}\mathrm{Li}\) reaction.

The drift gap was chosen to increase the neutron-induced signal amplitude by allowing the charged reaction products to deposit a substantial fraction of their energy in the gas volume. The simulation indicated that a \qty{10}{mm} drift gap is sufficient to contain the dominant alpha-particle energy deposition.

This geometry involves a trade-off between deposited-energy collection and spatial resolution.
A larger drift gap increases the neutron-induced deposited energy and can improve the separation from low-amplitude background through threshold selection, but it can also broaden the reconstructed charge distribution through transverse diffusion and the projected topology of the alpha and \(^{7}\mathrm{Li}\) tracks.

Finally, the simulation framework served as a theoretical baseline for understanding the detector's operational characteristics.
By comparing the simulated detection efficiency and pulse-height spectra with experimental data, we could cross-check the measured efficiency.
Further details regarding the simulation setup can be found in Ref.~\cite{Kim2025}.

\subsection{Detector Fabrication}
\label{subsec:fabrication}

\begin{figure}[t]
\centering
\includegraphics[width=0.5\linewidth]{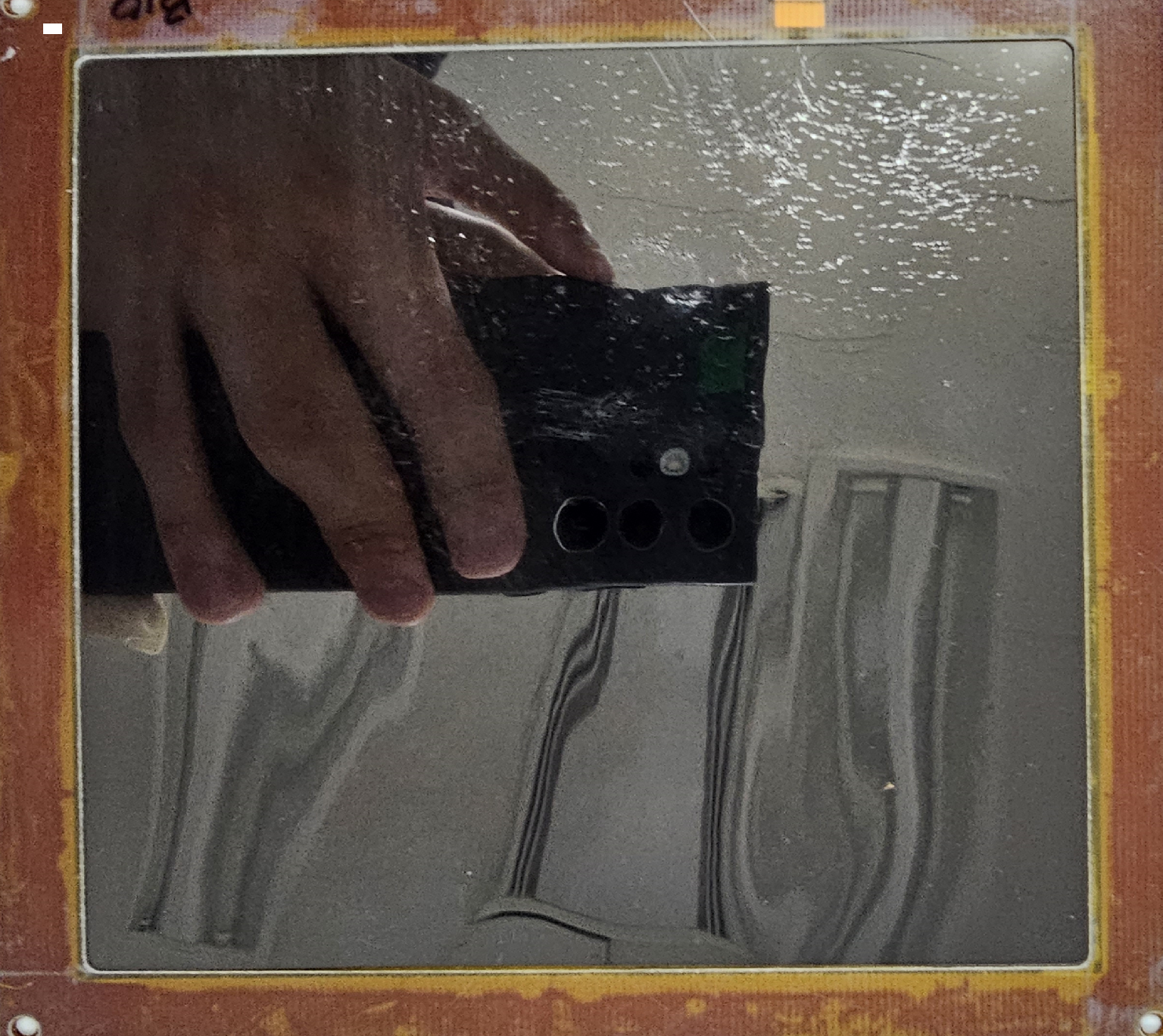}
\caption{Photograph of the cathode foil coated with \qty{1.5}{\micro\meter} of \({}^{10}\mathrm{B}_4\mathrm{C}\). The mirror-like surface finish indicates a metallic coating and the absence of obvious non-reflective impurities.}
\label{fig:cathode}
\end{figure}

Guided by the simulation results, we developed a BGEM detector to validate the design under cold-neutron irradiation.
For the neutron converter, a \qty{1.5}{\micro\meter} thick layer of \({}^{10}\mathrm{B}_4\mathrm{C}\) was coated onto the cathode.
Fig.~\ref{fig:cathode} shows the \({}^{10}\mathrm{B}_4\mathrm{C}\)-coated cathode.
Although previous thin-film studies suggest that a coating thickness of approximately \qty{3}{\micro\meter} can yield higher detection efficiency~\cite{Hoglund2012JAP_B4Cfilms}, we selected a thickness of \qty{1.5}{\micro\meter} to prioritize mechanical stability and mitigate the risk of film peeling or delamination.

Strict quality control was maintained during deposition to minimize boron nitride (\(\mathrm{BN}\)) formation arising from residual nitrogen in the chamber.
The formation of \(\mathrm{BN}\) impurities lowers the effective atomic density of \({}^{10}\mathrm{B}\) within the converter, reducing the macroscopic absorption cross-section and the resulting neutron capture probability.
The coating quality was primarily evaluated by visual inspection of the surface luster.
A high-quality \({}^{10}\mathrm{B}_4\mathrm{C}\) layer shows a characteristic metallic sheen, whereas \(\mathrm{BN}\)-contaminated coatings typically appear translucent or whitish.
All cathode foils used in this work exhibited a metallic sheen consistent with a \({}^{10}\mathrm{B}_4\mathrm{C}\)-rich coating.

The amplification stage employs \qtyproduct[product-units=single]{10 x 10}{cm^2} GEM foils, which were produced at the facility described in Ref.~\cite{ABBAS2023168723}.
Each foil consists of a \qty{50}{\micro\meter} thick polyimide substrate clad on both sides with \qty{5}{\micro\meter} thick copper layers.
The foils feature standard biconical holes with outer and inner diameters of \qty{70}{\micro\meter} and \qty{50}{\micro\meter}, respectively, arranged in a hexagonal pattern with a pitch of \qty{140}{\micro\meter}.

To verify that the detected signals originate from neutron capture reactions in the boron layer, we prepared two otherwise identical detector setups: one equipped with the \({}^{10}\mathrm{B}_4\mathrm{C}\)-coated cathode and a reference detector without the coating.
By comparing the response of these two detectors under the same beam conditions, we could confirm the neutron sensitivity attributed specifically to the \({}^{10}\mathrm{B}_4\mathrm{C}\) layer.

\subsection{DAQ system}
\label{subsec:daq}

The data-acquisition (DAQ) system was implemented using a custom single-board readout system for the 512 strip channels.
The readout plane consists of orthogonal X and Y strip electrodes with 256 strips in each coordinate.
Four APV25 ASICs~\cite{FRENCH2001359} were used for front-end signal processing, with two chips assigned to the X-axis strips and two chips assigned to the Y-axis strips.
The DAQ board is based on a System-on-Chip (SoC) FPGA platform with embedded Linux.
The board performs APV25 control, trigger handling, waveform buffering, event formatting, and data storage within a single compact system.
This architecture removes the need for multi-board trigger distribution or inter-board synchronization, while preserving strip-level charge information for all 512 channels.

The shaped trigger signal was delivered to the DAQ board as a NIM-format signal through a LEMO input.
The APV25 pipeline and the FPGA buffering logic allow samples corresponding to a programmable trigger latency to be retained and read out after a trigger.
For each accepted trigger, ADC values from all 512 channels were stored event by event, preserving the strip-level charge information for subsequent offline analysis.
The single-board implementation provides a compact readout solution for a 512-channel GEM detector without requiring multi-board synchronization.

\section{Experimental Setup and Method}
\label{sec:setup}

To evaluate the neutron detection performance of the fabricated prototype, beam tests were conducted at the Bio-REF cold neutron beamline of the HANARO (High-flux Advanced Neutron Application ReactOr) facility~\cite{CHOI2018236}.
During the experiment, the reactor operated at a thermal power of \qty{27}{MW}.
The Bio-REF beamline delivers a monochromatic cold neutron beam with a mean energy of approximately \qty{4.03}{meV} (\(\lambda=\qty{4.5}{\angstrom}\)) and a maximum neutron flux of \(4.8\times10^{6}~\mathrm{n\,cm^{-2}\,s^{-1}}\).
The incident beam profile was defined by a \(\mathrm{B}_4\mathrm{C}\) slit, resulting in a beam size of \qty{40}{mm} (horizontal) \(\times\) \qty{0.2}{mm} (vertical).

The detector was operated with an \(\mathrm{Ar}/\mathrm{CO}_{2}\) gas mixture in a 70/30 ratio.
The gas was supplied in an open-flow configuration at atmospheric pressure, with a flow rate of \qty{5}{L/h}.
This gas condition was used for both the efficiency and imaging measurements.
The drift gap between the \({}^{10}\mathrm{B}_{4}\mathrm{C}\)-coated cathode and the first GEM foil was \qty{10}{mm}.
This gap was selected to ensure sufficient energy deposition from the charged reaction products in the drift volume, rather than to optimize the intrinsic spatial resolution.

The trigger signal was derived from the bottom copper layer of the second GEM foil via AC coupling.
The signal was first processed by an ORTEC 142PC charge-sensitive preamplifier.
Subsequently, pulse shaping and amplification were performed using an ORTEC 474 timing filter amplifier to optimize the signal for triggering.

We performed a comparative study of the count rates from both the coated and uncoated detectors as a function of the applied voltage.
Measurements were acquired under both beam-on and beam-off conditions.
This approach allowed us to evaluate the dependence of neutron detection efficiency on the detector gain and to quantify the specific contribution of the boron coating to the total event rate.

This comparison was used to identify a voltage region in which the boron-coated detector shows a clear neutron-correlated excess while the uncoated reference remains close to the background level.

\subsection{Absolute Efficiency Measurement Setup}
\label{subsec:efficiency_setup}

The absolute detection efficiency was evaluated using a Ce:LiCAF scintillator as a reference detector.
Previous measurements using a \qty{2}{mm} thick Ce:LiCAF scintillator reported a relative efficiency of \(82\pm5\%\) compared with GS20 glass~\cite{IWANOWSKA2011319}.
In the present experiment, however, the reference crystal had a thickness of 1 cm.
The approximation of an effectively complete neutron absorber is therefore justified by the large absorption probability of 4.03 meV cold neutrons in the crystal. 
The \({}^{6}\mathrm{Li}\) number density of the crystal is approximately \(9.1\times10^{21}~\mathrm{cm^{-3}}\).

The reference-detector count rate was obtained after neutron/gamma separation using the pulse-shape difference of the Ce:LiCAF scintillation signal~\cite{IWANOWSKA2011319}.
In particular, a rise-time cut was applied to reject gamma-induced events, which have a pulse shape different from that of neutron-induced \({}^{6}\mathrm{Li}(n,\alpha)t\) events.
Only events passing the neutron-selection cut were used to determine \(R_{\mathrm{LiCAF,on}}\) and \(R_{\mathrm{LiCAF,off}}\).
This procedure suppresses gamma contamination in the reference count rate used for the BGEM efficiency calculation.
The 1 cm Ce:LiCAF crystal was therefore treated as an effectively complete absorber for 4.03 meV neutrons.

The efficiency of the boron-coated GEM (\(\epsilon_{\text{BGEM}}\)) was determined by comparing its background-subtracted count rate to that of the reference detector under identical beam conditions, as defined by Eq.~\ref{eq:efficiency}:
\begin{equation}
    \epsilon_{\text{BGEM}} = \frac{(R_{\text{BGEM,on}} - R_{\text{BGEM,off}})}{(R_{\text{LiCAF,on}} - R_{\text{LiCAF,off}})} \times \epsilon_{\text{LiCAF}},
    \label{eq:efficiency}
\end{equation}
where \(R_{\text{BGEM,on}}\) and \(R_{\text{BGEM,off}}\) denote the measured count rates of the boron-coated GEM detector under beam-on and beam-off conditions, respectively.
Similarly, \(R_{\text{LiCAF,on}}\) and \(R_{\text{LiCAF,off}}\) represent the corresponding count rates obtained from the reference Ce:LiCAF scintillator under the same experimental conditions.
The term \(\epsilon_{\mathrm{LiCAF}}\) denotes the effective neutron efficiency assigned to the reference detector after neutron-event selection.
In the present analysis, this value was approximated as unity based on the near-complete absorption probability of \qty{4.03}{meV} neutrons in the \qty{1}{cm} thick Ce:LiCAF crystal.

The coated/uncoated comparison was introduced not only to verify the neutron sensitivity associated with the \({}^{10}\mathrm{B}_{4}\mathrm{C}\) converter layer, but also to define a practical operating region.
Since switching off or changing the bias of the coated cathode would modify the electric-field configuration and charge-collection conditions, an uncoated detector with otherwise identical geometry provides a more direct control reference for separating the converter-induced neutron signal from beam-related and detector-related background contributions.

\section{Results}
\label{sec:results}

\begin{figure}[htbp]
    \centering
    \includegraphics[width=0.6\linewidth]{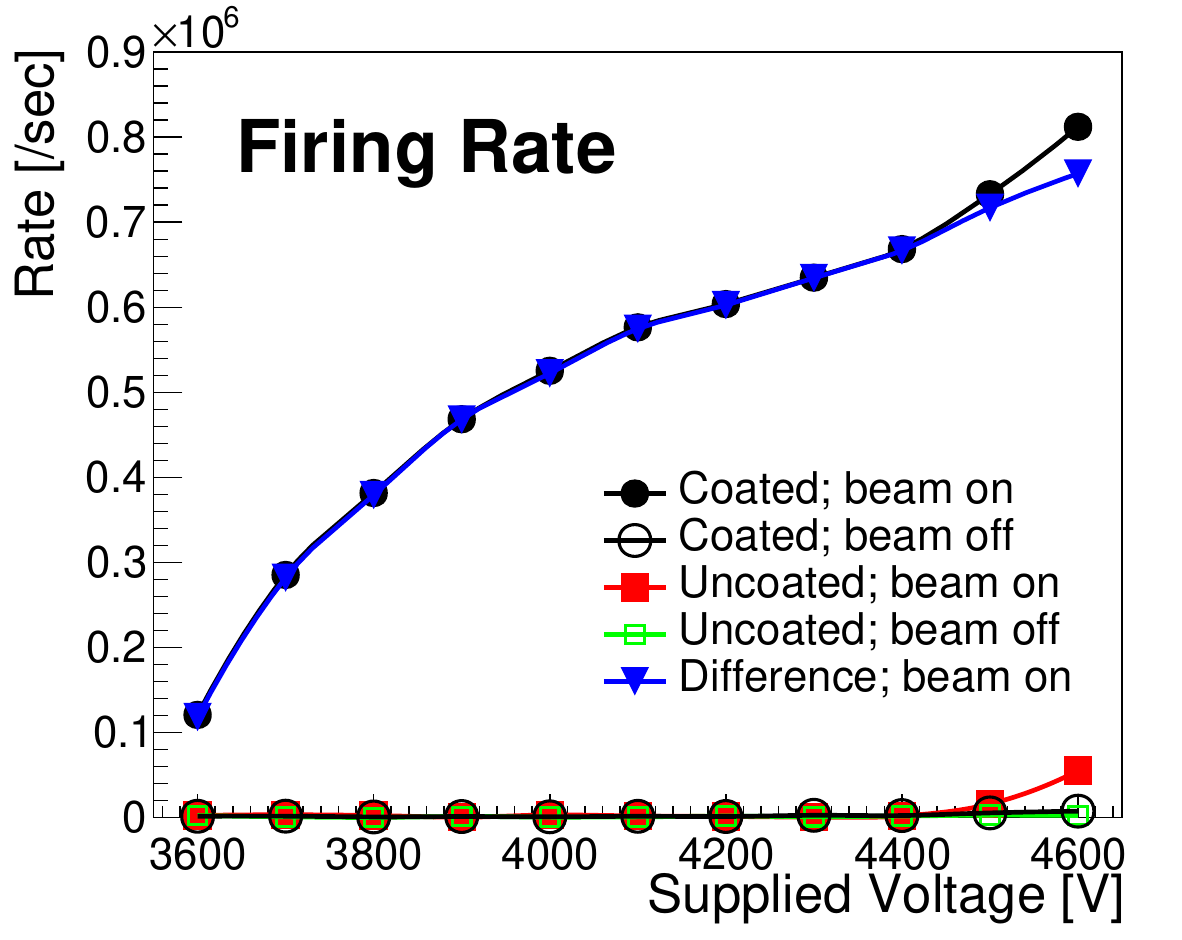}
    \caption{Measured event rates as a function of applied high voltage for both the boron-coated and uncoated detectors under beam-on and beam-off conditions. The curve labeled ``Difference'' represents the count rate of the coated detector minus that of the uncoated detector, illustrating the net neutron-capture signal.}
    \label{fig:rate-plot}
\end{figure}

Neutron-induced signals were first examined using event-rate measurements.
Fig.~\ref{fig:rate-plot} compares the count rates of the boron-coated and uncoated detectors as a function of the applied voltage under both beam-on and beam-off conditions.
In the beam-off state, both configurations yielded negligible background counts, demonstrating that the electronic noise level remains low despite the low discrimination threshold.
Under beam irradiation, only the boron-coated detector exhibited a substantial increase in event rate, characteristic of neutron capture signals.

At applied voltages above approximately \qty{4500}{V}, the uncoated reference detector exhibited a small beam-correlated increase in count rate. 
Since the origin of this component cannot be uniquely determined from the present measurement alone, we conservatively interpret it as a possible gamma-induced or beam-related background contribution. 
One plausible source is the \qty{478}{keV} prompt gamma ray associated with neutron capture in boron-containing beamline components, including the \(\mathrm{B}_4\mathrm{C}\) beam-defining slit. 
This residual response becomes noticeable mainly above the operating point selected for the efficiency measurement. 
At \qty{4400}{V} with a discriminator threshold of \qty{60}{mV}, the beam-off background was strongly suppressed, whereas a clear neutron-correlated excess was observed only for the boron-coated detector.

\begin{figure}[htbp]
    \centering
    \includegraphics[width=0.6\linewidth]{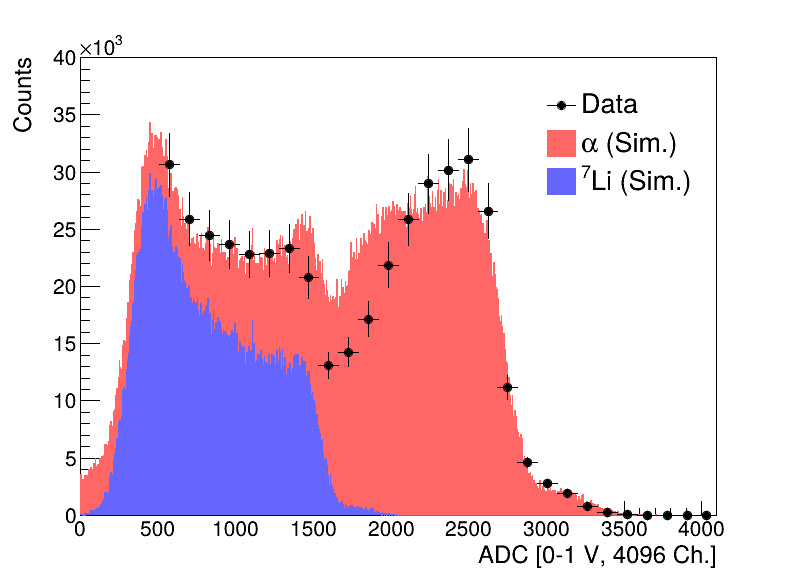}
    \caption{Pulse-height spectrum measured with the BGEM detector. The simulation results corresponding to the energy deposition of \({}^{7}\mathrm{Li}\) ions and \(\alpha\) particles are overlaid as stacked filled histograms.}
    \label{fig:mca}
\end{figure}

Neutron detection was further confirmed by analyzing the pulse-height spectrum.
\Cref{fig:mca} presents the measured pulse-height spectrum from the BGEM detector, superimposed with the simulation results.

A \qty{4}\percent~Gaussian smearing was applied phenomenologically to the simulated energy-deposition spectrum to represent the finite detector response.
The simulated energy-deposition spectrum was introduced phenomenologically to represent detector-response broadening, including gas-amplification fluctuations, electronics noise, and charge-collection variations.
The smeared spectrum was not used to extract the detection efficiency or any quantitative detector-response parameter.
It was used only as a qualitative consistency check to verify that the measured pulse-height distribution is compatible with the expected energy deposition from the \(\alpha\) and \({}^{7}\mathrm{Li}\) products of the \({}^{10}\mathrm{B}(n,\alpha){}^{7}\mathrm{Li}\) reaction.

\subsection{Detection Efficiency}
\label{subsec:Efficiency}

To distinguish neutron signals from electronic noise and gamma-ray background, the count rates were measured as a function of the discriminator threshold at a fixed operating voltage of \qty{4400}{V}.
As summarized in \Cref{tab:efficiency}correspond to the reduced-beam configuration used for the absolute-efficiency measurement.
The background count rate under beam-off conditions decreases rapidly in the low-threshold region, dropping from \qty{1265.0}{Hz} at \qty{30}{mV} to \qty{1.2}{Hz} at \qty{60}{mV}.
Above this value, the background rate becomes negligible.
Consequently, a threshold of \qty{60}{mV} was selected for the efficiency measurement, as it provides sufficient rejection of noise events while maintaining a significant net neutron count rate of \qty{200.0}{Hz}.

\begin{table}[htbp]
    \centering
    \caption{Rate measurement results at \qty{4400}{V}.}
    \label{tab:efficiency}
    \begin{tabular}{@{}lllll@{}}
    \toprule
    Voltage & Threshold & Beam off (Hz) & Beam on (Hz) & On - Off \\
    \midrule
    4400 V & 30 mV & 1265.0 & 1699.8 & 434.8 \\
    4400 V & 40 mV & 124.6 & 352.9 & 228.3 \\
    4400 V & 50 mV & 24.2 & 232.5 & 208.3 \\
    4400 V & 60 mV & 1.2 & 201.2 & 200.0 \\
    4400 V & 70 mV & 0.7 & 188.8 & 188.1 \\
    \bottomrule
    \end{tabular}
\end{table}

Based on this operating condition, the net count rates were measured to be \qty{200.0}{Hz} for the BGEM and \qty{2300}{Hz} for the reference LiCAF detector.
Using Eq.~\ref{eq:efficiency} and taking the cold-neutron absorption probability of the reference crystal to be effectively unity after neutron-event selection, the absolute cold-neutron detection efficiency of the BGEM was calculated to be \(8.69\pm0.20\%\).
The reported uncertainty corresponds to the statistical error only.
This experimental result is in agreement with the \GEANT simulation of the detector geometry, which predicted a detection efficiency of \qty{8.997}{\percent}~\cite{Kim2025}.

Systematic effects related to the reference-detector neutron-selection acceptance, threshold setting, detector alignment, beam-position reproducibility, dead time, and possible residual gamma contamination were not included.
Thus, the reported efficiency should be interpreted as the absolute efficiency measured under the selected operating condition, with a statistical uncertainty only.

\subsection{Imaging and Estimation of Spatial Resolution}
\label{subsec:spatial_resolution}

\begin{figure}[htbp]
  \centering
  \begin{subfigure}[t]{0.49\linewidth}
    \centering
    \includegraphics[width=\linewidth]{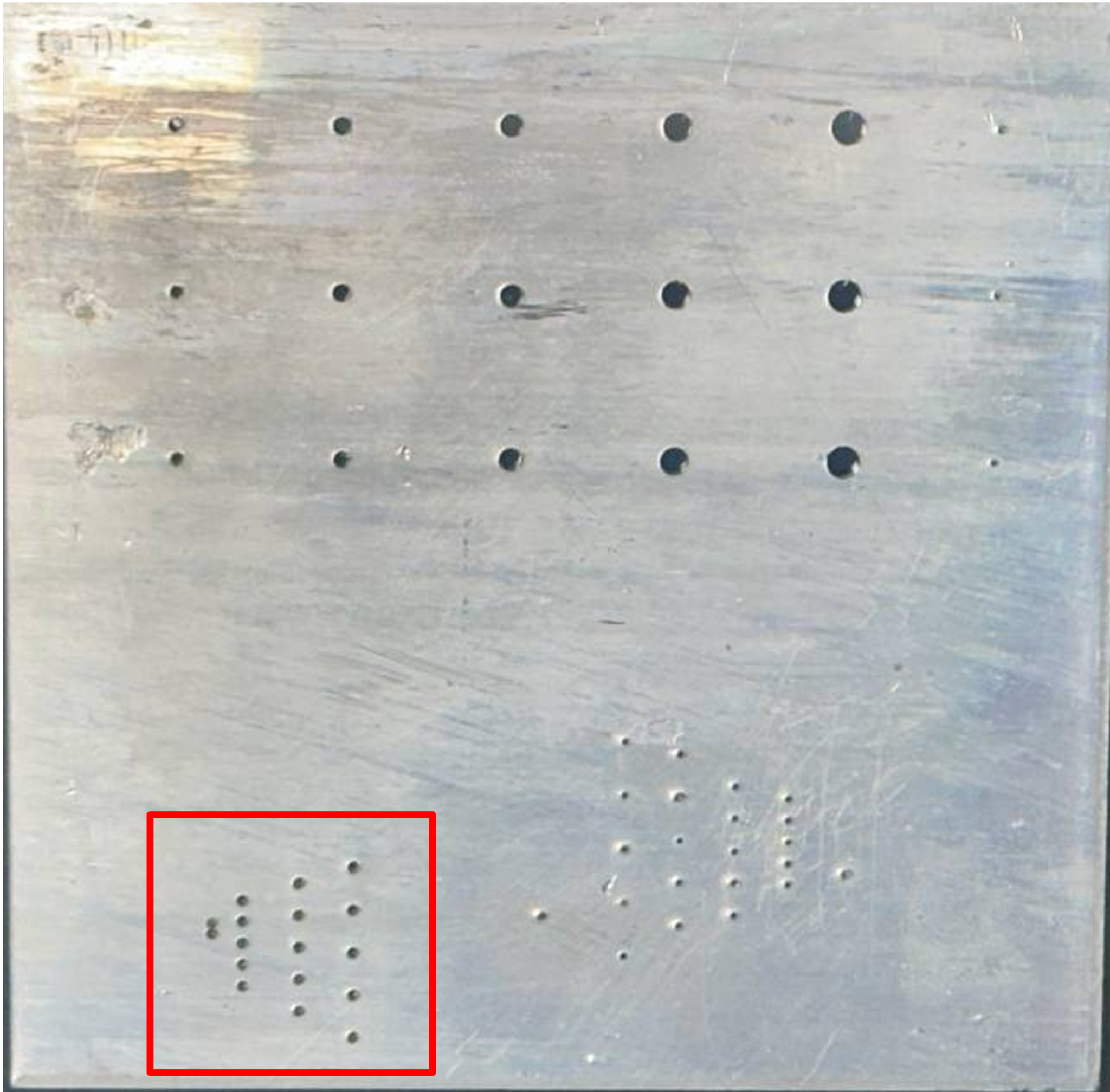}
    \caption{}
    \label{fig:cd_mask_photo}
  \end{subfigure}
  \hfill
  \begin{subfigure}[t]{0.49\linewidth}
    \centering
    \includegraphics[width=\linewidth]{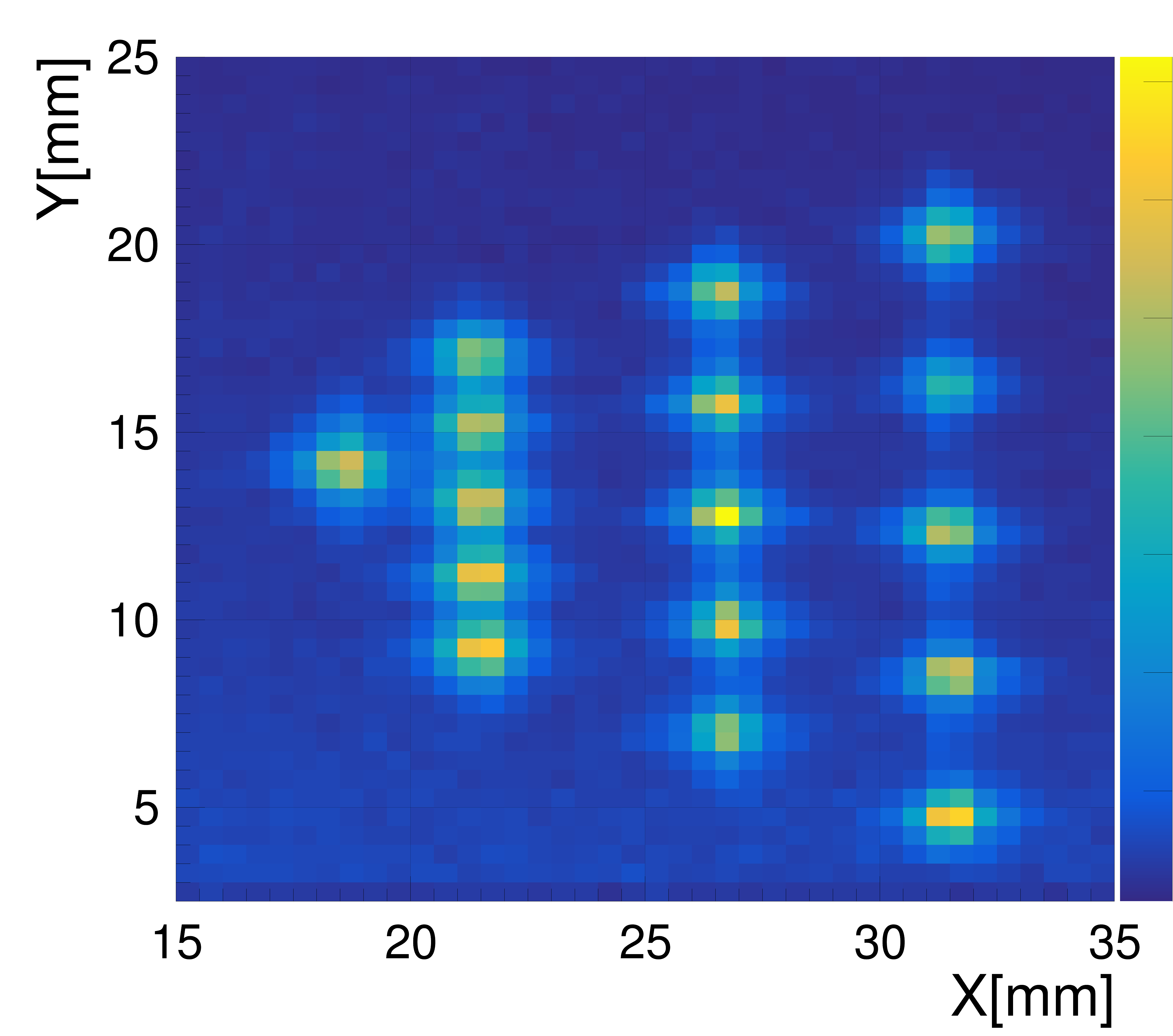}
    \caption{}
    \label{fig:holemap_zoom}
  \end{subfigure}
  \caption{Cd-mask imaging used for the imaging and spatial-resolution measurement. (a) Photograph of the Cd hole mask containing holes of several diameters. (b) Zoomed view of the reconstructed 2D hit-density map corresponding to the selected area.}
  \label{fig:cd_mask_imaging}
\end{figure}

The imaging capability of the BGEM detector was evaluated using a cadmium (Cd) mask with circular holes of different diameters.
Figure~\ref{fig:cd_mask_imaging} shows a photograph of the mask and the corresponding reconstructed hit distribution.
Clearly separated hole images are observed in the selected region, demonstrating the position-sensitive response of the detector.

For each triggered event, the strip charge was defined as the pedestal-subtracted peak ADC value of the APV25 waveform.
The hit position was reconstructed using a charge-weighted center-of-mass method,
\begin{equation}
    x_{\mathrm{CoM}}=\frac{\sum_i x_i q_{x,i}}{\sum_i q_{x,i}}, \qquad
    y_{\mathrm{CoM}}=\frac{\sum_j y_j q_{y,j}}{\sum_j q_{y,j}},
    \label{eq:com_reconstruction}
\end{equation}
where \(q_{x,i}\) and \(q_{y,j}\) are the strip charges in the \(x\)- and \(y\)-directions, respectively.
The reconstructed position therefore represents the centroid of the induced charge distribution on the readout plane, rather than the exact neutron-capture point in the converter.
Accordingly, the measured edge-spread width characterizes the reproducibility of the reconstructed centroid under the present imaging conditions, not the full physical length of individual \(\alpha\) or \({}^{7}\mathrm{Li}\) ionization tracks.

\begin{figure}[t]
    \centering
    \begin{subfigure}[t]{0.48\textwidth}
        \centering
        \includegraphics[width=\linewidth]{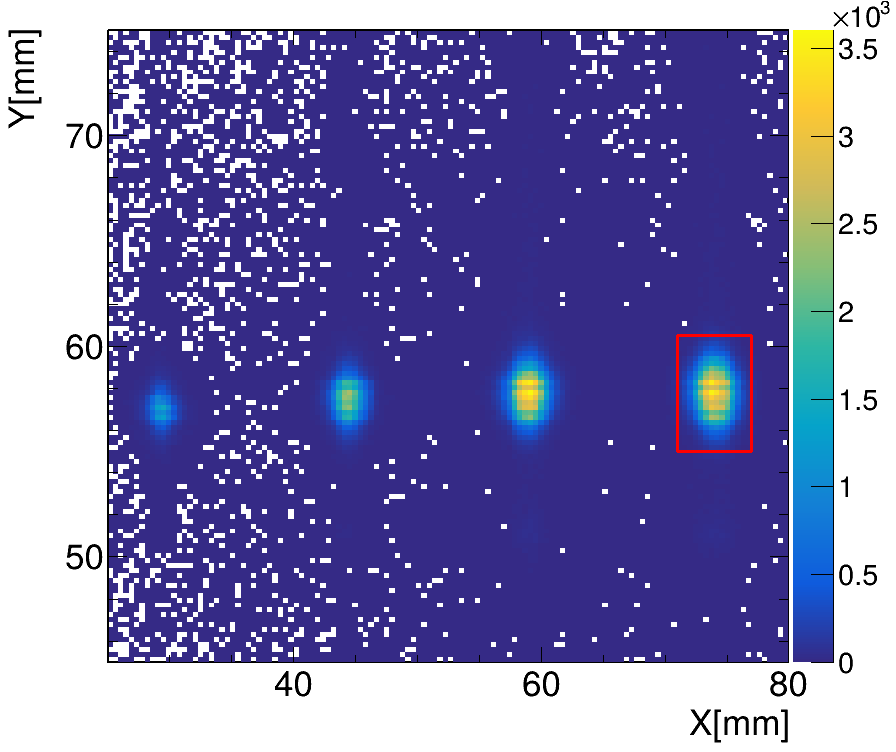}
        \caption{}
        \label{fig:radial_erfc_resolution_2d}
    \end{subfigure}
    \hfill
    \begin{subfigure}[t]{0.48\textwidth}
        \centering
        \includegraphics[width=\linewidth]{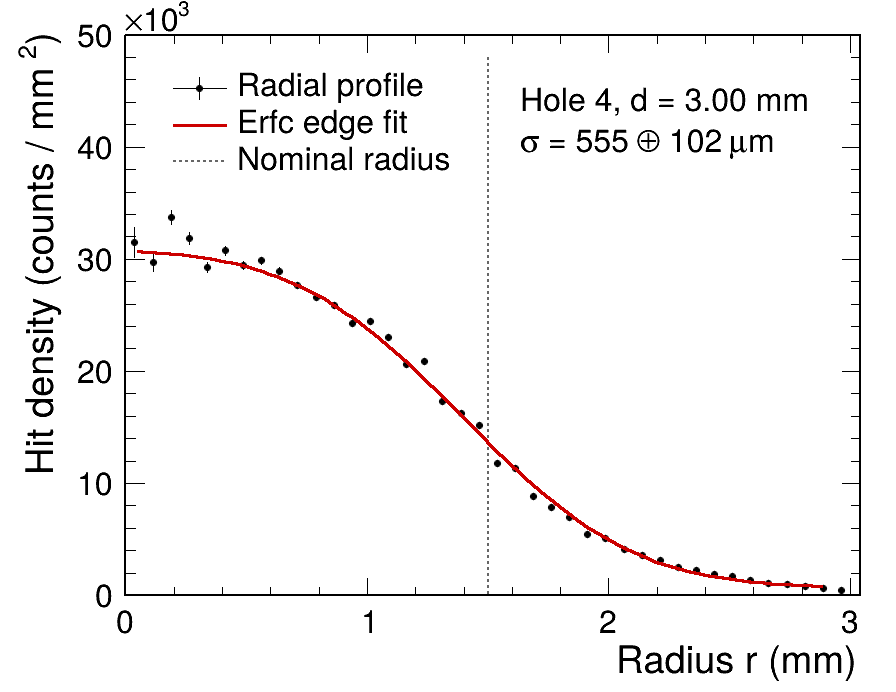}
        \caption{}
        \label{fig:radial_erfc_resolution_profile}
    \end{subfigure}
    \caption{
    Spatial-resolution estimation from the radial edge profile of a Cd-mask hole.
    (a) Reconstructed two-dimensional hit-density map, with the red rectangle indicating the region used for the analysis.
    (b) Radial hit-density profile for the selected 3.00~mm-diameter hole, fitted with a complementary error function.
    The dashed line marks the nominal hole radius of 1.50~mm, and the fitted Gaussian-equivalent edge-spread width is $\sigma = 555 \oplus  102~\mathrm{\mu m}$.}
    \label{fig:radial_erfc_resolution}
\end{figure}

Figure~\ref{fig:radial_erfc_resolution} shows the reconstructed two-dimensional hit-density map and the radial edge-spread analysis for a selected 3.00~mm-diameter hole.
Hits within the red rectangle in Fig.~\ref{fig:radial_erfc_resolution}(a) were transformed into radial coordinates relative to the fitted hole center, and the radial density was normalized by the corresponding annular area.
The edge profile was fitted with a complementary error function, yielding
\begin{equation}
    \sigma = 555 \oplus  102~\mathrm{\mu m}.
    \label{eq:edge_sigma_result}
\end{equation}
The result demonstrates that the reconstructed image edge is described by a sub-millimeter Gaussian-equivalent edge-spread width under the present cold-neutron beam conditions.

\section{Discussion}

The present work should be understood as a system-level cold-neutron beamline benchmark for a single-converter BGEM detector.
Compared with cathode-coated double-GEM detectors based on resistive-chain readout with only a few electronic channels, the full-strip APV25 readout records strip-level charge information event by event and provides a more direct basis for centroid-based image reconstruction.

\section{Conclusion}
\label{sec:conclusion}

We have developed and characterized a \({}^{10}\mathrm{B}\)-based double-GEM neutron detector featuring a \({}^{10}\mathrm{B}_{4}\mathrm{C}\)-coated cathode and an orthogonal strip readout with 256 strips per coordinate, corresponding to 512 readout channels in total.
The detector was tested at the HANARO Bio-REF cold-neutron beamline and demonstrated a clear beam-correlated neutron signal in the boron-coated configuration.

The absolute cold-neutron detection efficiency was measured to be \(\epsilon_{\mathrm{BGEM}} = (8.69 \pm 0.20)\%~\mathrm{(stat.)}\) for \qty{4.03}{meV} neutrons.
Cd-mask imaging yielded a Gaussian-equivalent edge-spread width of $\sigma = 555 \oplus 102~\mathrm{\mu m}$, which characterizes the reconstructed system-level image sharpness under the present beam and analysis conditions.
This work provides a practical cold-neutron benchmark for a full-strip APV25-based single-converter BGEM architecture.
The results provide a reference point for future optimization of converter thickness, drift geometry, and readout strategy for neutron beam-monitoring and imaging applications.

\section*{Acknowledgments}
This work was supported by the 2024 Research Fund of the University of Seoul (I. Park).
This research was also supported by the National Research Foundation of Korea (NRF) grant funded by the Korea government (MSIT) (No. 2017R1A2B4006644) and by the Basic Science Research Program through the NRF funded by the Ministry of Education (No. 2018R1A6A1A06024977).
We gratefully acknowledge Dr. Sangjin Jo and Dr. Myungguk Moon of the Korea Atomic Energy Research Institute (KAERI) for their technical assistance with the \(\mathrm{B}_{4}\mathrm{C}\) coating trials on the GEM and cathode foils and for providing the neutron beam facility for this study.

\bibliographystyle{unsrt}
\bibliography{ref.bib}

\end{document}